\documentclass[aps,showpacs,twocolumn,superscriptaddress]{revtex4}
\usepackage{graphicx}
\begin{document}

% ----->   TITLE   <-----

\title{Scalar Field Dark Matter: non-spherical collapse and late time 
behavior}

% ----->   AUTHORS   <-----

\author{Argelia Bernal}
\affiliation{Departamento de F{\'\i}sica, Centro de Investigaci\'on y de
              Estudios Avanzados del IPN, A.P. 14-740, 07000 M\'exico
              D.F.,MEXICO.}

\author{F.~Siddhartha~Guzm\'an}
\affiliation{Instituto de F\'{\i}sica y Matem\'{a}ticas, Universidad
              Michoacana de San Nicol\'as de Hidalgo. Edificio C-3, Cd.
              Universitaria, A. P. 2-82, 58040 Morelia, Michoac\'{a}n,
              M\'{e}xico.}

% --->   DATE

\date{\today}

% ----->   ABSTRACT   <-----

\begin{abstract}
We show the evolution of non-spherically symmetric balls of a 
self-gravitating scalar field in the Newtonian regime or equivalently an 
ideal self-gravitating condensed Bose gas. In order to do so, we use a 
finite differencing approximation of the Shcr\"odinger-Poisson 
(SP) system of equations with axial symmetry in cylindrical coordinates. 
Our results indicate: 1) that spherically symmetric ground state 
equilibrium configurations are stable against non-spherical 
perturbations and 2) that such configurations 
of the SP system are late-time attractors for non-spherically symmetric 
initial profiles of the scalar field, which is a generalization of such 
behavior for spherically symmetric initial profiles. Our system and the 
boundary conditions used, work as a model of scalar field dark matter 
collapse after the turnaround point. In such case, we have found that the 
scalar field overdensities tolerate non-spherical contributions to the 
profile of the initial fluctuation.
\end{abstract}

% ----->   PACS
                                                                                

% ----->   MAKETITLE   <-----

\maketitle

% ----------------------
% ----->   BODY   <-----
% ----------------------

% ----->     INTRODUCTION

\section{Introduction}
\label{sec:introduction}

Recently, scalar fields have played different roles in several scenarios 
related to astrophysical phenomena. The reason is that such fields are 
quite common in theoretical physics, specially branches related to 
theories beyond the standard model of particles, high dimensional theories 
of gravity an tensor-scalar theories alternative to General Relativity. In 
the present research we deal with the scalar field dark matter model 
(SFDM), which assumes the dark matter to be a classical minimally coupled 
real scalar field determined by a cosh-like potential. Such potential 
behaves exponentially at early stages of the universe and as a free field 
(quadratic potential) at late times, which provides the field with the 
necessary properties to mimic the behavior and successes of cold dark 
matter at cosmic scales. In fact in 
\cite{Sahni2000,MatosUrena2000,MatosUrena2001} 
it was shown that the mass parameter of the scalar field gets fixed by a 
desired cut-off of the power spectrum, which has two effects: i) the 
theory gets fixed and ii) there is no overabundance of substructure, 
which standard cold dark matter cannot achieve. One important consequence 
is that the boson has to be ultralight with masses around $m \sim 
10^{-21,-23}$eV. This is a substantially important bound, because in the 
standard dark matter models there are no such ultralight dark matter 
candidates. The benefit obtained however, is two fold: the scalar field 
can represent a Bose Condensate of such ultralight particles and the 
Compton wavelength forbids the scalar field to form cuspy structures. 
In fact, in \cite{FerrerGrifols2001} it was shown that in order for the 
scalar field interaction to become long range, the system only needs 
to be condensed.

After the fluctuation analysis about this candidate and its corresponding
concordance with observations -the model mimics the properties of the
$\Lambda$CDM at cosmic scales-, the next step has to be in the direction
of the study of structure formation and the explanation of local
phenomena, like rotation curves in galaxies. Fortunately there have been
important advances in such direction 
\cite{Arbey2001,Mbelek2004}. About the gravitational 
collapse, in \cite{AlcubierreDark2002} it was shown that relativistic 
self-gravitating scalar field configurations can be formed when they have 
galactic masses provided the mass of the boson is ultralight. 
Nevertheless, because the gravitational field in galaxies is weak, the 
race turned into the newtonian limit of the system of equations, which was 
developed in \cite{GuzmanUrena2003}. The price to be paid is that it is 
not possible to apply the approach at very early stages of the evolution 
of the universe, and the profit is that the scalar field in the 
non-relativistic regime provides a clear interpretation within the Bose 
Condensate formalism and classical Quantum Mechanics. In both cases, the 
strong gravity and the Newtonian regimes, a wide range of arbitrary 
spherically symmetric initial configurations collapse and form 
gravitationally bounded and virialized objects with a smooth density 
everywhere called oscillatons (except those that are related to unstable 
initial configurations that collapse into black holes in the strong field 
regime) \cite{GuzmanUrena2004,AlcubierreDark2003}. This property seems to 
be fundamental in order to form galactic halos, because several high 
resolution observations are consistent with regular galactic dark matter 
profiles in the center of the galaxies \cite{flatcore}, which implies that 
this type of condensate could be an alternative to solve the cuspy density 
profiles of dark halos.

Two pieces of the model that are in progress are the high energy fully 
relativistic case, including exponential-like scalar field potentials and 
the free field weak energy newtonian, both are complementary and necessary 
to explore the SFDM hypothesis. Inspired in a late-time astrophysical 
scenario, at stages after the turnaround point where weak field applies 
and the field is free, the question is whether dark matter halos
are gravitationally bounded objects of scalar field  which have been
formed through a gravitational collapse of initial scalar field
overdensities. The newtonian version of the Einstein-Klein-Gordon system
of equations is provided by the Schr\"odinger-Poisson equations (SP),
which are the ones that lead the gravitational collapse of the system.
This approximation should work for the evolution of an initial density
profile after the epoch when the overdensity fluctuation starts to evolve
independently of the cosmic expansion.

In the recent past, it has been found that in spherical symmetry the SP
system has equilibrium solutions of two types: stable, for which the 
wave function is nodeless (called sometimes ground state 
configurations) and others, for which the wave function has nodes 
(called sometimes excited configurations) that decay into ground state 
solutions. From these solutions only the ground state ones are
stable \cite{GuzmanUrena2004,Harrison2002}; even further, it has been 
found that such configurations behave as late-time attractors for 
initially quite arbitrary spherically symmetric density profiles 
\cite{GuzmanUrena2003,GuzmanUrena2006}. 

In \cite{GuzmanUrena2003} it was shown that free scalar field 
overdensities after the turnaround virialize and tend to form 
ground state configurations. Due to a very general scale 
invariance of the Schr\"odinger-Poisson system of equations such results 
were also valid for a structure of arbitrary mass 
\cite{GuzmanUrena2004}. In \cite{GuzmanUrena2006} was shown that the 
spherical collapse of SFDM tolerates the introduction of a 
self-interaction term, which on the other hand is associated to the 
self-interaction term of a self-gravitating Bose-condensate 
\cite{Wang2001}, and helps at allowing diverse sizes and masses of 
the final configurations.

In this paper we go a step forward and study the collapsing process of 
non-spherical initial configurations, in particular those involving at 
most quadrupolar terms. In order to achieve this goals an axisymmetric 
code that solves the SP system is needed. We choose to deal with 
cylindrical coordinates denoted by $(x,z)$, where $x$ is the 
radial coordinate and $z$ the axial coordinate. The SP system for the free 
field case in these coordinates reads

\begin{eqnarray}
i \frac{\partial \psi}{\partial t} &=& -\frac{1}{2} \left( 
				\frac{\partial^{2} \psi}{\partial x^2}
	                      + \frac{1}{x}\frac{\partial \psi}{\partial x}
                              + \frac{\partial^{2} \psi}{\partial z^2}
				\right)
		    + U \psi \label{eq:schroedinger}\\
\frac{\partial^{2} U}{\partial x^2} &+&
\frac{1}{x}\frac{\partial U}{\partial x} +
\frac{\partial^{2} U}{\partial z^2} = 
\psi^{\ast}\psi.\label{eq:poisson}
\end{eqnarray}

\noindent where $\hbar = c = 1$ and we are using the rescaled 
variables $x\rightarrow mx$, $z\rightarrow mz$, $t \rightarrow mt$ and 
the wave function $\psi \rightarrow \sqrt{4\pi G}\psi$. This set of 
coupled partial differential equations in two spatial dimensions plus 
time is the core of the present manuscript. As mentioned before, these 
equations have stable solutions in spherical symmetry (see 
\cite{GuzmanUrena2006} for solutions also including a non-linear term 
in the Schr\"odinger equation); such solutions have shown to be not only 
stable, but also late time attractors for quite arbitrary initial 
density profiles \cite{GuzmanUrena2004,GuzmanUrena2006}. Therefore, our 
main task in the present manuscript will be to show that such solutions 
are still attractors even for non spherically symmetric initial density 
profiles.

In the next section we present the code we constructed for 
the present purpose. In section \ref{sec:nonsphericalperturbations} we 
show how a ground state configuration reacts under non-spherical 
perturbations. In section \ref{sec:nonsphericalcollapse} we show the 
evolution and fate of non-spherical initial density profiles. Finally 
in section \ref{sec:conclusions} we draw some conclusions.

% ----->     THE CODE

\section{The code}
\label{sec:code}

% ->     Description

\subsection{Description}

The present code is built under the same numerical finite differences 
method used in the spherically symmetric case in 
\cite{GuzmanUrena2004,GuzmanUrena2006}. We approximate the continuous 
equations 
(\ref{eq:schroedinger}-\ref{eq:poisson}) using centered finite 
differencing for both coordinates $x$ and $z$ on a uniform grid defined by 
$x=p\Delta x$ and $z=q\Delta z$, $p,q$ integers; we used the same 
resolution in both directions ($\Delta x = \Delta z$). The spatial 
differential operator is the same in both equations and we dealt with 
both in the same manner: aside of the usual finite differencing 
expression for the space derivatives only two delicate items were included 
related to the first order derivative with respect to $x$ in 
(\ref{eq:schroedinger}-\ref{eq:poisson}): i) we staggered the grid in the 
$x$-direction in order to avoid the divergence of such term and ii) we 
transformed such term into $\frac{1}{x}\frac{\partial \psi}{\partial x} = 
2 \frac{\partial \psi}{\partial x^2}$, with the last expression a 
derivative with respect to $x^2$.

{\it Schr\"odinger equation.} In the present case this is the evolution 
equation of the system. We discretize time $t = n \Delta t$, $n$ an 
integer and $\Delta t$ the resolution in time. We solve this equation 
using a second order accurate explicit time integrator, which is a modified 
version of the usual three steps iterative Crank-Nicholson method
\cite{Teukolsky2001}.

Instead of using a characteristic analysis of the propagating 
modes to set an open boundary at the edges of the domain, we decided to 
use a sponge in the outermost region of the domain. The sponge is a 
concept used successfully in the past when dealing with the Schr\"odinger 
equation (for detailed analyses see \cite{GuzmanUrena2004,Israeli1981}). 
This technique consists in adding up to the potential 
in the Schr\"odinger equation an imaginary potential. The result is that 
in the region where this takes place there is a sink of particles, and 
therefore the density of probability approaching this region will be 
damped out, with which we get the effects of a physically open 
boundary.

{\it Poisson equation.} Equation (\ref{eq:poisson}) is an elliptic 
equation for $U$ which we solve using the 2D five-point stencil for the 
derivatives and a successive over-relaxation (SOR) iterative algorithm 
with optimal acceleration parameter (see e.g.\cite{Smith1965} for details 
about SOR). In order to impose boundary conditions we made sure the 
boundaries were far enough for the mass $M=\int |\psi|^2 
d^3x$ to be the same along the three faces of the domain and used the 
monopolar term of the gravitational field; that is, we used the value 
$U = -M/r$ along the boundaries with $r=\sqrt{x^2+z^2}$ for the 
gravitational potential. At the axis we demanded the gravitational 
potential to be symmetric with respect to the axis.

% -> Spherical Initial Data

\subsection{Construction of equilibrium initial data}
\label{subsec:Construction of equilibrium initial data}

{\it Initial data.} The nature of the SP system allows one to have plenty 
of freedom about choosing the initial data, that is: once we choose an 
initial wave function $\psi$ we integrate (\ref{eq:poisson}) at initial 
time; this means that the initial wave function is quite arbitrary. In 
fact we implement this type of initial data for the case of 
non-spherical collapse. Nevertheless, for the purpose of testing our 
numerical techniques, we decided to use initial data corresponding to 
spherically symmetric ground state configurations. Here we briefly 
describe how these data are obtained.

In spherical symmetry equations (\ref{eq:schroedinger},\ref{eq:poisson}) 
read:

\begin{eqnarray}
i\frac{\partial}{\partial t} \psi &=& 
	-\frac{1}{2r} \frac{\partial^2}{\partial r^2} (r\psi) 
	+ U \psi
\label{eq:sph_schroedinger}\\
\frac{\partial^2}{\partial r^2} (rU) &=& r \psi \psi^\ast . 
\label{eq:sph_poisson}
\end{eqnarray}

\noindent where $r=\sqrt{x^2+z^2}$ is the spherical radial coordinate. 
It is assumed a time dependence of the type 
$\psi = \phi(r) e^{i \omega t}$, and demand the conditions of 
regularity at the origin $\phi(0)=\partial_x \phi(0)=0$ and isolation 
$\phi(x\rightarrow \infty) = 0$, the system becomes an eigenvalue 
problem where the frequency of the wave function is the eigenvalue. The 
system to be solved reads

\begin{eqnarray}
\frac{\partial^{2}}{\partial r^2}(r\phi)&=& 2 r (U-\omega) 
%+ 2\Lambda|\phi|^2\phi \, , 
\label{eq:sph_schroedinger_eq} \\
\frac{\partial^{2}}{\partial r^2}(rU)&=& r\phi^2 \, . 
\label{eq:sph_poisson_eq}
\end{eqnarray}

\noindent We use a shooting method that bisects the value of $\omega$ 
for a given central value of $\phi$ that satisfies the boundary 
conditions. That is, one constructs a one parameter family of solutions 
labeled by central field, and a given frequency is found for each value 
of the label as shown in \cite{GuzmanUrena2004,GuzmanUrena2006}. Excited 
solutions can be also constructed by allowing $\phi$ to vanish at a 
given number of points, but always demanding the satisfaction of the 
boundary conditions. Up to here the construction in spherical symmetry. 
Once we account with these data: i) we interpolated the wave function of 
the spherical data in the $xz$-grid and ii) resolved the 
Poisson equation (\ref{eq:poisson}), then we have initial data for 
ground state configurations in our axially symmetric domain.

The system (\ref{eq:sph_schroedinger}-\ref{eq:sph_poisson}) is invariant 
under a clever scaling property given by

\begin{eqnarray}
\{t,r,U,\phi\}
&\rightarrow &
\{
\lambda^{-2}\hat{t},\lambda^{-1}\hat{r},\lambda^{2}\hat{U},\lambda^2\hat{\phi}
\}
        \label{scaling1}\\
\{\rho, M, K, W \}
&\rightarrow&
\{
\lambda^4\hat{\rho}, \lambda \hat M, \lambda^3 \hat K,\lambda^3 \hat W
\}
        \label{scaling2}
\end{eqnarray}

\noindent where $U$ is the gravitational potential, $\phi$ is the 
spatial part of the wave function, $\rho$ is the density of probability, 
$M$ is the integral of $\rho$, $K$ and $W$ are the expectation value of 
the kinetic and gravitational energy respectively, and $\lambda$ is 
a scaling parameter. Property (\ref{scaling1}-\ref{scaling2}) implies 
that if a solution is found for a given central field value 
$\hat{\phi}(0)=\hat{\phi}_0$ (e.g. $\hat{\phi}(0) = 1 \Rightarrow 
\rho(0) = 1$) it is possible to build the whole branch of 
ground state equilibrium 
configurations. For instance, if the plot $M ~ vs ~ \rho(0)$ is to be 
constructed, we know from \cite{GuzmanUrena2004} that for 
$\hat{\phi}(0)=\hat{\rho}(0) = 1$ we have $\hat{M} = 2.0622$; 
using the relations (\ref{scaling1}-\ref{scaling2}) for $\phi$ and $M$ we 
find $\lambda = \left(\rho/\hat{\rho} \right)^{1/4} = M/\hat{M}$, which 
implies that the desired plot is given by the function $M = 2.0622 
(\rho(0))^{1/4}$ for all central values of the scalar field density. This 
function is used later on when showing the attractor behavior of these 
configurations.

% ----->     Tests

\subsection{Testing the code with ground state configurations}

The steps followed in the construction of the solutions to equations 
(\ref{eq:schroedinger}-\ref{eq:poisson}) can be summarized 
as follows: 1) choose a type of initial data for the $t=0$ slice 
$\psi({\bf x},0)$, 2) populate the $xz$ grid with those data, 3) solve 
equation (\ref{eq:poisson}) and get a gravitational potential, 4) using 
such potential leap the system using (\ref{eq:schroedinger}) a $\Delta t$ 
time slice further, 5) use the obtained wave function to solve again 
(\ref{eq:poisson}), 6) then repeat the loop 4, 5 either until the physics 
starts going wrong (physical quantities lose convergence, dissipative 
effects show up) or the cpu-time cannot be afforded.

We test this code using systems whose properties we already know from 
their construction, these are the ground state equilibrium 
configurations constructed in the previous subsection; they show a 
particular property: in the continuum 
limit the wave function oscillates with constant frequency $\omega$, which 
implies the density of probability $\rho = |\psi|^2$ and therefore the 
gravitational potential $U$ remain time-independent. Unfortunately we do 
not account with infinite resolution and therefore we are solving the 
discretized versions of equations 
(\ref{eq:schroedinger},\ref{eq:poisson}); instead we solve 
the second order finite differencing approximation of those equations, 
which is nothing but a truncated expansion of the solution of the 
functions involved. Let us describe the situation with an example: 
assume we start with a ground state configuration with $\psi({\bf 
0},0)=1$ and therefore the central density $\rho({\bf 0},t)=\rho({\bf 
0},0)=1$ at all times; first, as we are solving a truncated system of 
equations the time-independence of the central density cannot be 
satisfied in a strict fashion, instead we can at most demand the central 
density to converge to 1 at all times; second, due to the truncation 
error of the finite differencing we are perturbing the system all the 
times and therefore the system should behave as an stable 
configuration that is perturbed and thus should oscillate with the modes 
obtained from a perturbation theory analysis. Therefore we have a 
two-fold trap to verify whether or not the code is solving the physical 
system whose properties we know beforehand.

\begin{figure}[htp]
\includegraphics[width=4cm]{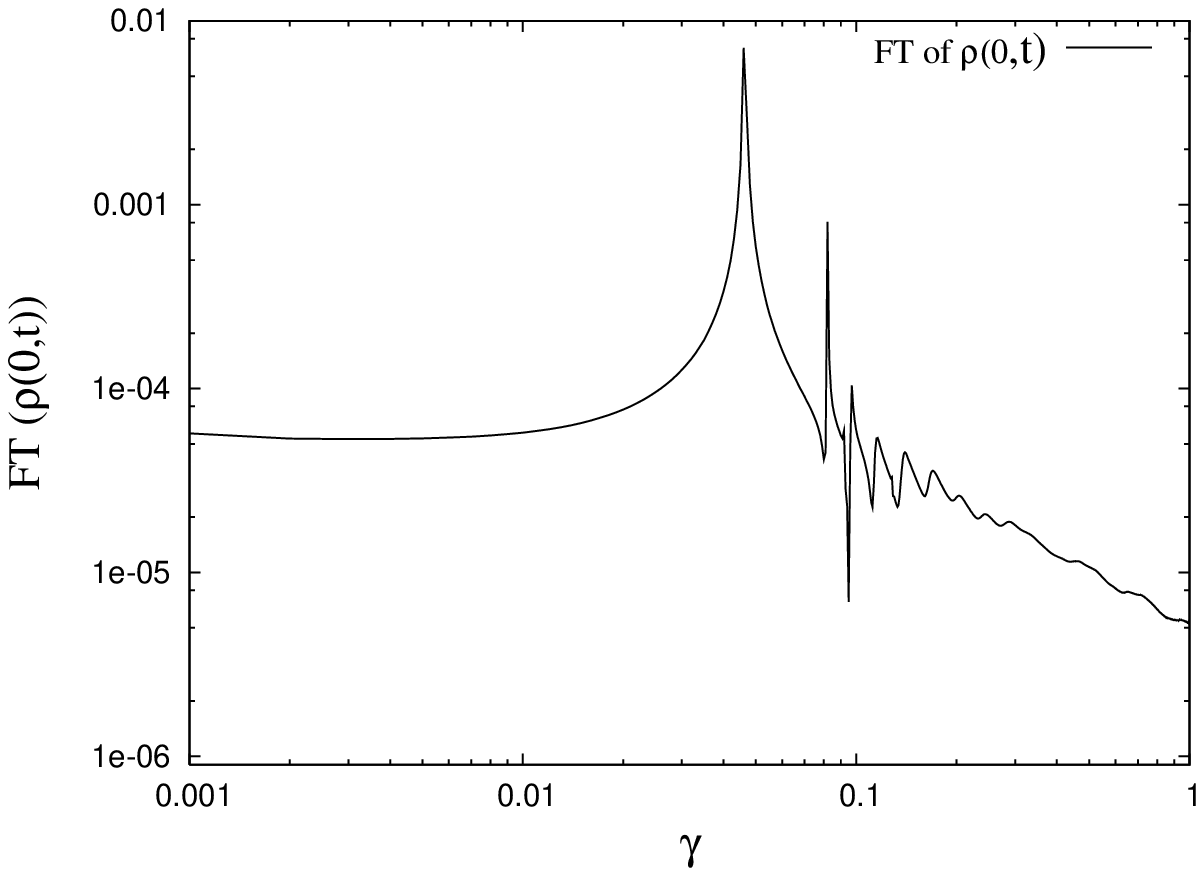}
\includegraphics[width=4cm]{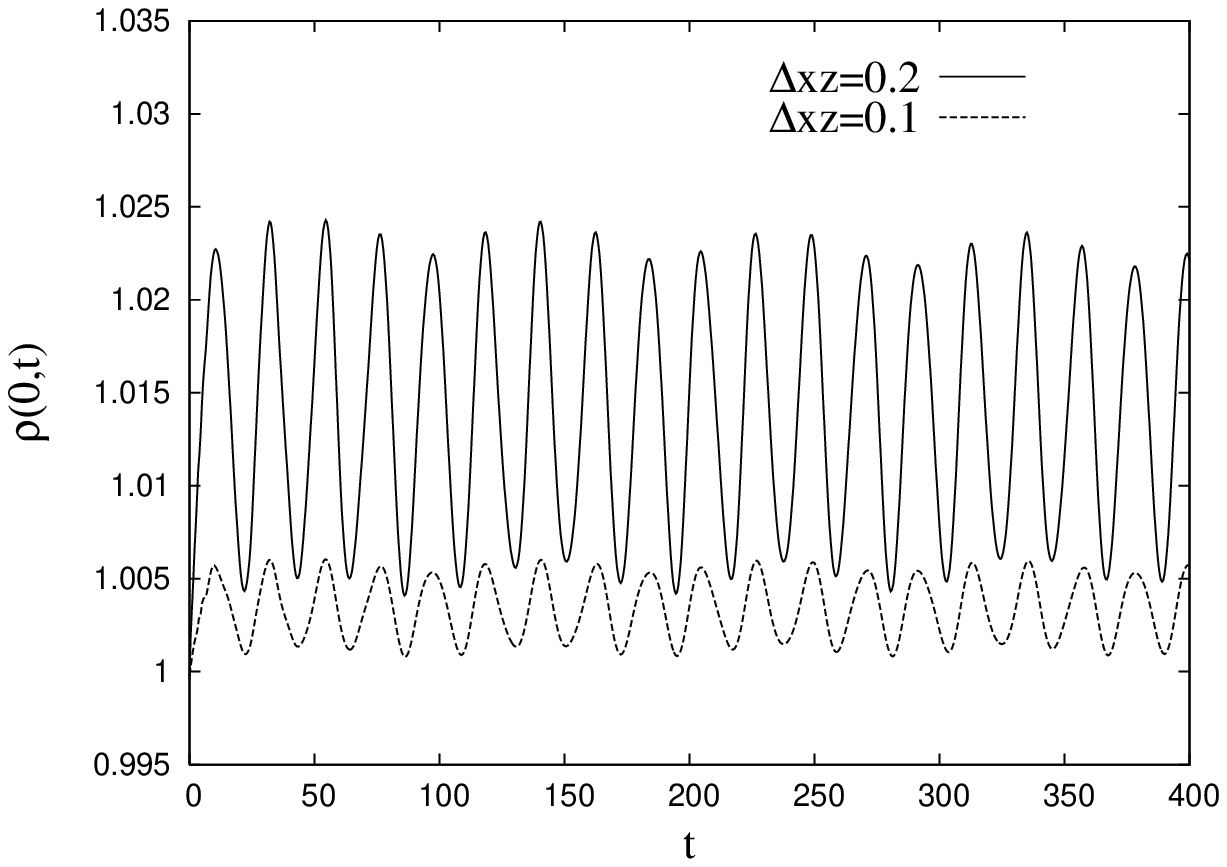}
\caption{\label{fig:dft} Left: we show the Fourier transform 
of the central density of the configuration after the evolution has 
been performed; the main peak shows up at $\gamma=0.046$, which 
coincides with the result found using 
the perturbation theory in spherical symmetry from \cite{GuzmanUrena2004}. 
Right: the second order convergence of the central density to the value 
one is shown using two different resolutions. The runs were carried out on 
a $x \in [0,20],~ z\in [-20,20]$ domain with resolutions $\Delta x = 
\Delta z =0.1, 0.2$. A Cauchy type convergence test decides whether or not we 
have convergence: the fact that the low resolution run shows a central 
density four times bigger than the one with the double resolution with 
respect to the value one, indicates the second order convergence to 
one.} 
\end{figure}

With all this in mind we evolved such a  configuration with 
$\psi({\bf 0},0)=1$ with our code and the results are as follows. The 
Fourier transform in Fig. \ref{fig:dft} of the central density 
reveals that the main frequency of oscillation is $\gamma=0.046$ which 
coincides with the result predicted by the first order radial perturbation 
theory developed in \cite{GuzmanUrena2004}; this indicates that despite 
the non spherical nature of our grid, the perturbation due to the 
discretization of the physical domain is spherical. On the right hand side 
of Fig. \ref{fig:dft} we show the second order convergence of the 
central density to one, which indicates that our approximations work as 
they should when we refine the grid and approach the continuum limit.

\begin{figure}[htp]
\includegraphics[width=7cm]{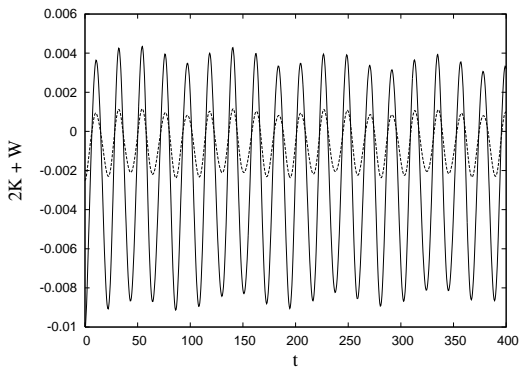}
\caption{\label{fig:sph_equilibrium_cooling} In this plot we show the 
meaning of virialized configurations. That is, due to discretization 
errors when solving the SP equations, there is an intrinsic error in the 
calculations. We say the system is virialized only in 
the continuum limit. This is the reason why we show here the second order 
convergence also for the virial relation. Because this example 
corresponds to an ground state equilibrium configuration, which we know 
is virialized, we can be confident that not only the evolution code, but 
also the diagnostics tools work fine.} 
\end{figure}

In Fig. \ref{fig:sph_equilibrium_cooling} we verify that 
ground state equilibrium configurations are virialized in the continuum 
limit, where the relation $2K + W = 0$ is satisfied with second order 
convergence. The quantities $K$ and $W$ are calculated as follows:

\begin{eqnarray}
K &=& -\frac{1}{2}\int \psi^{\ast} \nabla^2 \psi d^3x\\
W &=& \frac{1}{2}\int \psi^{\ast}U\psi d^3x
\end{eqnarray}

\noindent where the integrations are performed over the numerical 
domain.\\

% ----->     Non spherical perturbations

\section{Non-spherical perturbations}
\label{sec:nonsphericalperturbations}

It is still possible to use the discretization error to perturb a 
ground state equilibrium configuration in a non-spherical way, for 
instance, using 
different resolutions in the $x$ and $z$ directions (see \cite{Guzman2004} 
for the use of such trick in relativistic boson stars). However this 
time we choose to fully-truly perturb the system with the addition of 
a shell of particles as done in \cite{Seidel98}, but this time using 
non-spherical shells like in \cite{Balakrishna2006}. We start with a 
spherically symmetric ground state equilibrium configuration and add 
up a contribution proportional to a given spherical harmonic. Thus, 
we propose initial data given by

\begin{equation}
\psi({\bf x},t) = \psi_{ground} + \delta \psi
\label{eq:superposition}
\end{equation}

\noindent where $\delta\psi = e^{-i k_r 
\sqrt{x^2+y^2}}\left(\sum^{2}_{l=0} 
a_l Y^{0}_{l} \right)$, where the coefficients $a_l = A_l 
e^{-(\sqrt{x^2+z^2}-r_0)^2/\sigma^2}$ are gaussian like shells 
added to the real part of the wave function of the 
ground state configuration; on the other hand, the factor $e^{-i k_r 
\sqrt{x^2+y^2}}$ contributes with a non-zero initial speed of the 
perturbation; notice that the up labels of the spherical harmonics are 
all zero because otherwise they cannot be defined within an axially 
symmetric grid. Thus the complete set of parameters characterizing the 
perturbation are: $r_0$, $A_l$, $\sigma$ and $k_r$.

\begin{figure}[htp]
\includegraphics[width=7cm]{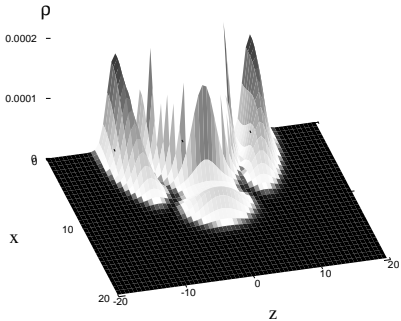}
\caption{\label{fig:initial-perturbed-rho} The initial density profile 
of the perturbed equilibrium configuration. The central blob corresponds 
to the equilibrium configuration, whose magnitude is one at the origin. 
The perturbation thus consists of two blobs coming with speed $k_r$ from 
the poles and a belt of particles over the equatorial plane.} 
\end{figure}

We practiced several combinations of coefficients $a_l$, including 
spherical perturbations containing only the $Y^{0}_{0}$ contribution and 
found similar results. We only present the one with the parameters: 
$r_0=10.0$, $A_0=A_1=0$, $A_2=0.02$, $\sigma=2.0$, $k_r=-2.0$ carried out 
on a grid $x \in [0,20],~ z\in[-20,20]$ with resolution $\Delta x = \Delta z 
=0.2$. The 
perturbation shows a quadrupolar contribution. In Fig. 
\ref{fig:initial-perturbed-rho} we show the density of probability 
$\rho$ at initial time. The mass of the shell is 0.8\% that of the 
ground state equilibrium configuration.

\begin{figure}[htp]
\includegraphics[width=4cm]{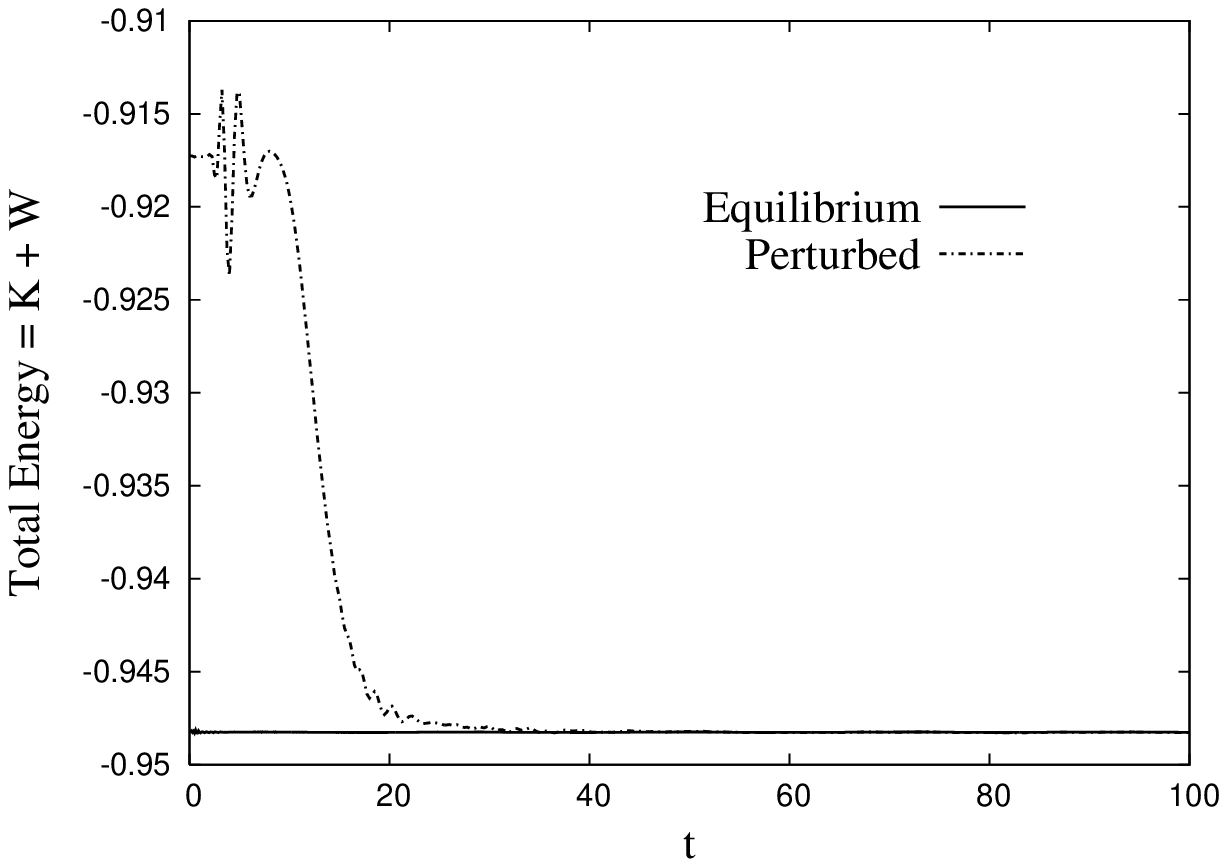}
\includegraphics[width=4cm]{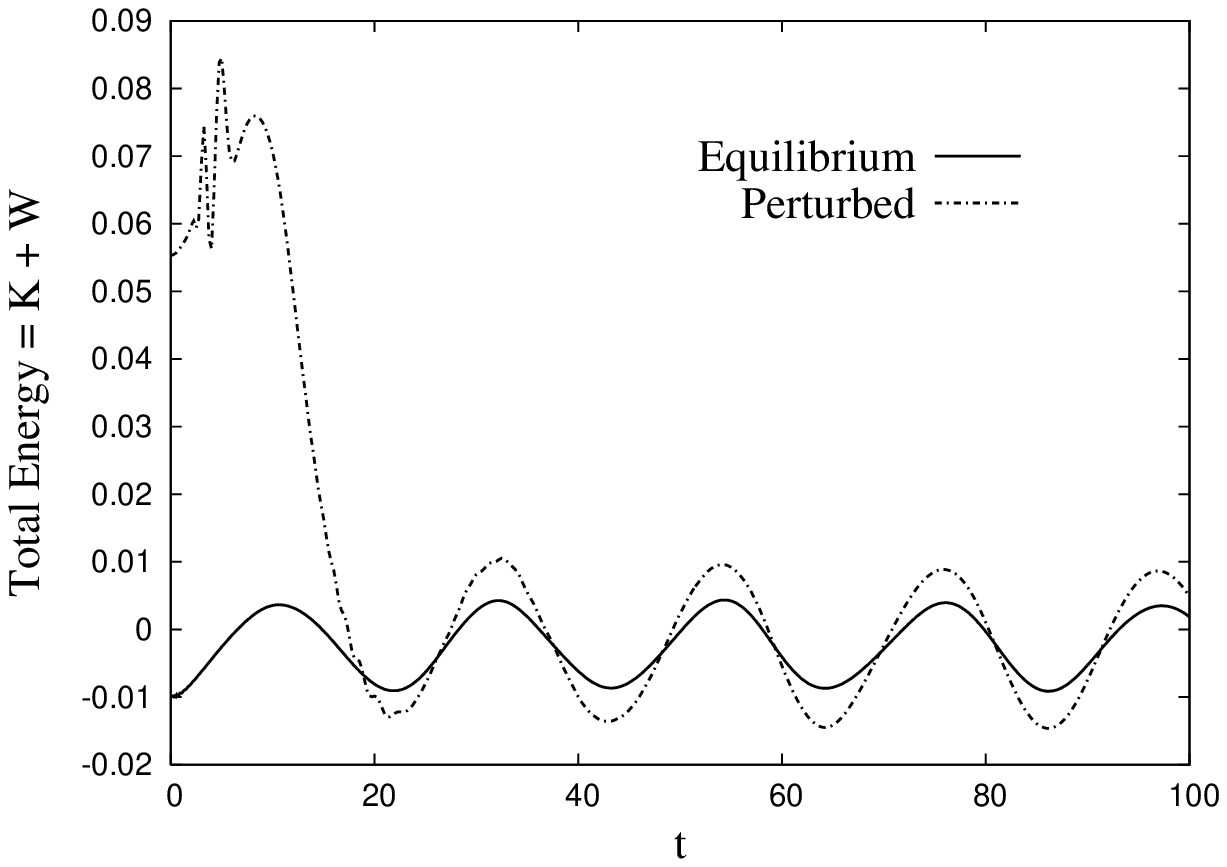}
\caption{\label{fig:perturbed-evolution} Left: evolution of the total 
energy; it can be observed that initially the whole system appears 
overwarmed due to the presence of the perturbation and its dynamical 
state; as the time runs, in a rather short time, the system recovers the 
total energy of the original configuration. Right: the 
quantity $2K+W$ is monitored and it relaxes and converges to zero with 
second order.}
\end{figure}

In Fig. \ref{fig:perturbed-evolution} we show the evolution of such 
system. What can be seen is that the system 
relaxes and virializes around a spherical ground state equilibrium 
configuration. The 
original equilibrium configuration is recovered, which was 
verified in the following terms: the mass $M$ and the total 
energy $E=K+W$ approach the values of the equilibrium configuration, the 
relation $2K+W$ converges to zero in the continuum limit after a short 
time, the system has a non-spherical initial shape and after a while the 
ellipticity converges to zero in the continuum limit. This reasons 
indicate that these ground state equilibrium configurations are stable 
against non-spherical 
perturbations that involve the introduction of a quadrupolar 
shell of particles.

% ----->     Non spherical collapse

\section{Non-spherical collapse}
\label{sec:nonsphericalcollapse}

In this section we essay a step forward in the obvious direction, that is, 
the collapse of non-spherical initial profiles for the SP system of 
equations. In fact we show that spherically symmetric ground states,
are late time attractors for initial configurations which are not
spherically symmetric.

% --> Requirements for a Non-spherical Collapse
 
\subsection{Requirements for a Non-spherical Collapse }

The method followed to verify that non-spherical initial 
configurations evolve toward a spherically symmetric ground 
state is as follows. 1) Given the evolution of a non-spherical initial 
profile $\psi({\bf x},t)$ is carried out, we obtain via a Fourier 
Transform of a physical quantity, e.g. the central density 
$\rho_c^{\psi}$, the frequency $\gamma$ of the fundamental mode of 
oscillation of the system; We assume that $\gamma$ corresponds to the 
characteristic frequency of oscillation of a linearly perturbed ground 
state $\psi_{perturbed}$ toward which $\psi$ is evolving to. 
This is because the intrinsic perturbation of the system associated to 
the discretization of the equations, we do not expect that, even for large 
evolution times, $\psi$ matches exactly a stationary ground state. 
2) Once we estimate $f$ (as done for the construction of Fig. 
\ref{fig:dft}) we can calculate the rescaling parameter $\lambda$ 
that relates $\psi_{perturbed}$ and the ground state with 
$\hat \psi({\bf x},0)=1$. Using the definition of frequency and the 
rescaling relation for $t$ given by (\ref{scaling1}) the rescaling 
parameter is calculated as $\lambda=(f/\hat f)^{1/2}$. 3) The rescaling 
parameter $\lambda$ lets us calculate the non-hat quantities in 
(\ref{scaling2}) for the ground state $\psi_{perturbed}$. Because we are 
not in the continuum limit we can at most demand that the physical 
quantities for $\psi({\bf x},t)$, such as its density $\rho^{\psi}_c$ and 
mass $M_{\psi}$, converge to those of a ground state configuration. 
4) However, the information related to the 
convergence to an equilibrium configuration is not enough and we also 
verify that the fate of $\psi$ is a spherical configuration. We define the 
ellipticity of the system as the integrated difference between 
$\rho_z^{\psi}=\rho(0,z,t)$ and $\rho_x^{\psi}=\rho(x,0,t)$ measured from 
the center of mass of the configuration. We observe for all our 
non-spherical initial configurations that after a transient period their 
ellipticity of the system relaxes and becomes zero in the continuum limit. 
5) Finally, for all the evolved initial configurations we verify that the 
virialization condition is satisfied as shown in the perturbation case in 
Sec. \ref{sec:nonsphericalperturbations}.

% --> Non-small perturbations

\subsection{Bigger perturbations}
\label{subsec:Non-Spherical and Non-Linear perturbation}

In order to illustrate this we show the evolution of 
a ground state $\psi_{ground}$ plus a considerable non-spherical density 
contribution which can be already considered to be a non-spherical initial 
profile. We evolved initial configurations of the form

\begin{equation}
\psi({\bf x},t)= \psi_{ground} + \Sigma_{l=0}^2 a_l Y^0_l
\end{equation}

\noindent where $a_l=A_l \exp[-r^2/\sigma_l^2]$ is a gaussian with 
$r=\sqrt{x^2+z^2}$ centered at the origin, width $\sigma_l$ and real 
amplitude $A_l$. Several runs were made for different $A_l$ and 
$\sigma_l$, and here we present only a representative one. In all cases 
the result is that the whole system evolves toward a ground 
state with a mass considerably bigger than that of $\psi_{ground}$
alone. The way in which the system evolves to a ground state 
is through the gravitational cooling process 
\cite{GuzmanUrena2004,GuzmanUrena2006}, which is powered by the ejection 
of scalar field. 

\begin{figure}[htp]
\includegraphics[width=7cm,angle=0]{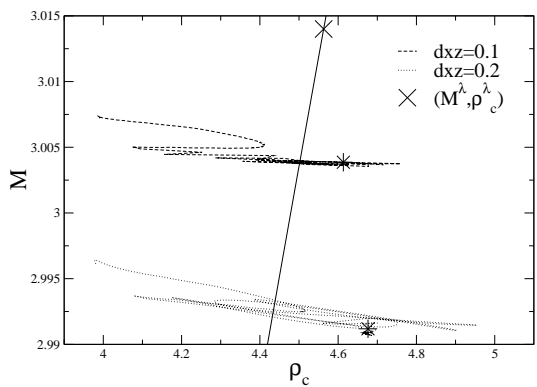}
\caption{\label{fig:EqplusYl1nonspherical-evolutionM} 
Evolution of the mass of the initial configuration versus the central 
density specified in the text. The resolutions used are $\Delta x 
= \Delta z=0.1$ 
(dotted) and $\Delta x = \Delta z=0.15$ (dashed). The solid line 
represents the 
branch of spherical ground states. considering our calculations are 
second order convergent, from these two runs with the respective 
resolutions, we infer that the configuration in the cuntinuum limit is 
that marked with a cross.} 
\end{figure}

This attractor behavior has been shown for spherically symmetric 
configurations in \cite{GuzmanUrena2006} and is  shown here for the first 
time for the case of non-spherical initial profiles. Here we present the 
results for the initial configuration for which $\psi_{ground}({\bf 0},0)=2$, 
$l=1$, $A_l=1$ and $\sigma_l=2$. The mass added to the system is the
order $\sim 3\%$ that of the original ground state configuration, which 
implies the 
following result: some of the added particles joined the ground state 
configuration and evolved toward another rescalled ground state configuration 
different from the original one, simply because it is not a small 
perturbation this time. The evolution was carried out on a grid 
$x \in [0,12],~ z\in[-12,12]$ with resolutions $\Delta x = \Delta z=0.1$ and 
$\Delta x = \Delta z=0.15$. For this configuration we found the 
fundamental frequency at 
late time to be $f=0.0983$, which gives a rescaling parameter of 
$\lambda=1.462$. The central density and mass of the ground state 
$\psi_{\lambda}$ associated with this specific $\lambda$ are 
$M_{\lambda}=3.014$ and $\rho_c^{\lambda}=4.563$. In Fig. 
\ref{fig:EqplusYl1nonspherical-evolutionM}, we show the evolution of the 
mass $M_{\psi}$ versus the central density $\rho_c^{\psi}$ of the state 
$\psi$ for the two different resolutions previously specified. The solid 
line is the branch of all the ground states constructed as indicated in 
section \ref{sec:code} and the cross symbol corresponds to the state 
with the parameters above. 

\begin{figure}[htp]
\includegraphics[width=7cm,angle=0]{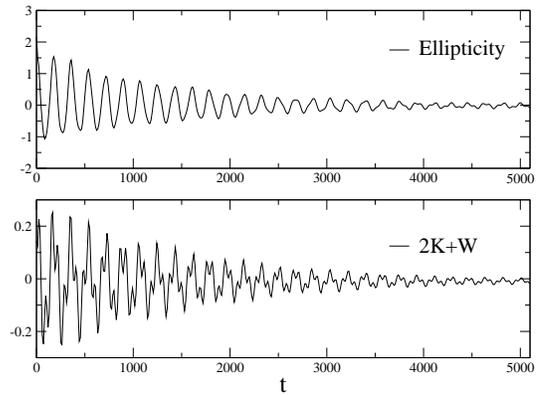}
\caption{\label{fig:EqplusYl1nonspherical-evolution} 
Evolution of the Ellipticity and the expression $2K+W$. It is shown that 
after a while the initial axisymmetric configuration evolves toward a 
spherical one as the Ellipticity goes to zero. On the 
other hand as  $2K+W$ oscillates around zero with decreasing amplitude we 
conclude that the system tends to a virialized state.} 
\end{figure}

Convergence of $M_{\psi}$ and $\rho_c^{\psi}$ quantities to the 
star in the plot is shown also in Fig. 
\ref{fig:EqplusYl1nonspherical-evolutionM}. The convergence is second 
order and the stars indicate the configuration we expect the system 
will relax onto for each resolution. A convergence analysis would reveal 
second order approach toward the configuration marked with the star. In 
Fig. \ref{fig:EqplusYl1nonspherical-evolution}, we show the ellipticity of 
the system; we can see that the system relaxes and becomes spherical. Also 
in this figure it is shown that the system is virialized. 

% --> Arbitrary Non-Spherical initial data
 
\subsection{Non-Spherical initial profiles}
\label{subsec:Arbitrary Non-Spherical initial data} 

\begin{figure}[htp]
\includegraphics[width=7cm,angle=0]{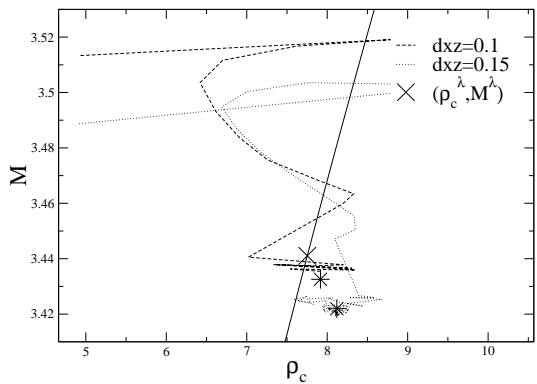}
\caption{\label{fig:Arbitrarynonspherical-evolutionM} 
Evolution of the central density and mass of the initial 
configuration specified in the text. We used once again resolutions 
$\Delta x = \Delta z=0.1$ (dashed) and $\Delta x = \Delta z=0.15$ 
(dotted). The 
solid line 
represents the branch of spherical ground states. As in the previous 
subsection, the stars correspond to the states our runs tend to, and the 
cross indicates the configuration we would achieve with infinite 
resolution.} 
\end{figure}

\begin{figure}[htp]
\includegraphics[width=7cm,angle=0]{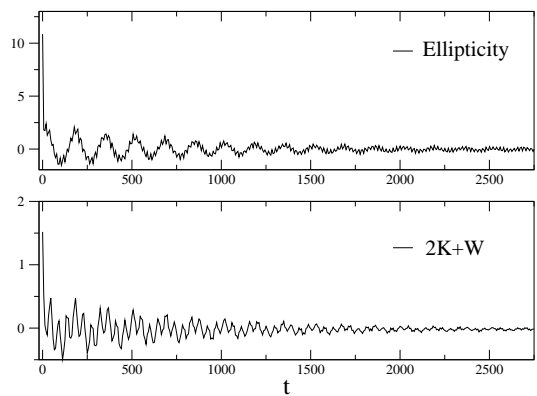}
\caption{\label{fig:Arbitratynonspherical-evolutionE} 
Temporal evolution of the Ellipticity and the expression $2K+W$ of the 
initial configuration specified in the text. It is shown that after a 
while the initial axisymmetric configuration evolves toward a spherical 
one as the Ellipticity goes to zero. On the other hand as  $2K+W$ 
oscillates around zero with decreasing amplitude we conclude that the 
system tends to a virialized state.} 
\end{figure}

What is in turn is to investigate the evolution of axisymmetric 
initial configurations. We choose these initial data to have the form

\begin{equation}
\psi({\bf x},t)= \Sigma_{l=0}^3 A_l \exp[-r^2/\sigma_l^2]Y^0_l
\end{equation}

\noindent where $r$ is as before, $\sigma_l$ is the width of a gaussian 
and $A_l$ a real amplitude. Several runs were made for different $A_l$ and 
$\sigma_l$ here we present the results for the initial configuration with 
$A_0=9.0$, $A_1=A_2=A_3=1.0$ and 
$\sigma_0=\sigma_1=\sigma_2=\sigma_3=1.5$. The evolution 
was carried out on a grid $x \in [0,12],~z\in [-12,12]$ with resolution 
$\Delta x = \Delta z=0.1$ and $\Delta x = \Delta z=0.15$. For this 
configuration 
we found a 
fundamental frequency $f=0.128$ that implies a rescaling parameter 
$\lambda=1.669$. The central density and mass  
of the ground state $\phi_{\lambda}$ associated with this specific 
$\lambda$ are $M_{\lambda}=3.441$ and $\rho_c^{\lambda}=7.75$ 
respectively, a point which is marked with a cross in Fig. 
\ref{fig:Arbitrarynonspherical-evolutionM}. In such 
Figure we show the evolution of the mass $M_{\psi}$ versus the central 
density $\rho_c^{\psi}$ of the state $\psi$ for the two grid resolutions 
previously specified. The solid line is the branch of all the ground 
states constructed as indicated in section \ref{sec:code}. The stars are 
the configurations the system is approaching to for the different 
resolutions; because we know such systems are only approximately an 
equilibrium configuration we practiced a Richardson extrapolation 
calculation and found that in terms of mass and central density, in the 
continuum limit -assuming the second order convergence we have in our 
algorithms- the cross is the configuration these results converge to.
Thus in Figs. \ref{fig:Arbitrarynonspherical-evolutionM} and 
\ref{fig:Arbitratynonspherical-evolutionE} we show that the system evolves 
toward a spherical and virialized configuration.

% ----->     CONCLUSIONS

\section{Conclusions}
\label{sec:conclusions}

We have presented a new code designed to solve the SP system with axial 
symmetry in cylindrical coordinates. We have shown it passes the 
necessary testbeds of stability and consistency with expected results 
related to spherically symmetric ground state equilibrium 
configurations.

We perturbed ground state equilibrium configurations with rather general 
axially symmetric shells of particles, which were proportional to the 
first axisymmetric spherical harmonics. Moreover, these shells were 
endowed with an initial speed toward the equilibrium configuration. The 
main result is that ground state configurations are stable against this 
type of perturbations.

Finally, we have evolved axisymmetric initial profiles, and showed that 
spherically symmetric ground state equilibrium configurations play the 
role of late-time attractors. This is a generalization of the same result 
when the initial profiles are spherically symmetric 
\cite{GuzmanUrena2006}. We have thus shown that the final state of a 
non-spherical self-gravitating scalar field fluctuation is a ground 
state spherically symmetric solution. Thus we showed that such  
configurations are late-time attractors not only for initial spherically 
symmetric configurations but also for quite general initial axisymmetric 
shapes.

In the context of the scalar field dark matter model we have quite a new
result: the collapse of overdensities tolerates an initial non-spherical 
contribution to the initial profiles, and moreover, initially axisymmetric 
profiles tend toward a spherical ground state.

About the process of virialization, with the results at hand we are unable 
to establish whether the gravitational cooling (powered by the emission of 
scalar field) is the only responsible for the relaxation of the system. 
The cases involving a non-zero quadrupolar contribution to the 
gravitational potential might allow one to formulate the study of another 
channel, that of the emission of gravitational waves, which would be the 
subject of a further investigation.

Definitely a major application of the tools shown in this paper is the
demonstration of the stability properties of rotating solutions to the
SP system of equations constructed in the past 
\cite{Harrison2002,Schupp1996}. It remains unclear whether excited state 
solutions representing spinning configurations are stable, and whether 
such solutions could contribute to the SFDM possibilities in terms of quick 
virialization of collapsed structures and other dynamical properties, for 
instance, the non-cupsy density profile might demand a considerably important
spherical contribution in the case the wave function has odd parity.

% ----->     ACKNOWLEDGMENTS

\acknowledgments

This research is partly supported by 
grants PROMEP UMICH-PTC-121 and CIC-UMSNH-4.9. The runs 
were carried out in the Ek-bek cluster of the ``Laboratorio de 
Superc\'omputo Astrof\'{\i}sico (LASUMA)'' at CINVESTAV-IPN. A. B. 
acknowledges a scholarship from CONACyT.

\end{document}